\documentclass[twocolumn,preprintnumbers,amsmath,amssymb,aps,prl,superscriptaddress]{revtex4}
\usepackage{graphicx}
\usepackage{times}

\begin{document}

\title{Disorder-perturbed Landau levels in high electron mobility epitaxial graphene}

\author{Simon Ma\"ero}
\affiliation{Laboratoire Pierre Aigrain, Ecole Normale Sup\'erieure\\
CNRS (UMR 8551), Universit\'e Pierre \& Marie Curie, Universit\'e Paris Diderot\\
24 rue Lhomond, 75231 Paris Cedex 05, France}

\author{Abderrezak Torche}
\affiliation{Laboratoire Pierre Aigrain, Ecole Normale Sup\'erieure\\
CNRS (UMR 8551), Universit\'e Pierre \& Marie Curie, Universit\'e Paris Diderot\\
24 rue Lhomond, 75231 Paris Cedex 05, France}

\author{Thanyanan Phuphachong}
\affiliation{Laboratoire Pierre Aigrain, Ecole Normale Sup\'erieure\\
CNRS (UMR 8551), Universit\'e Pierre \& Marie Curie, Universit\'e Paris Diderot\\
24 rue Lhomond, 75231 Paris Cedex 05, France}

\author{Emiliano Pallecchi}
\affiliation{Institut d'Electronique, de Micro\'electronique et de Nanotechnologie, CNRS (UMR 8520)\\
BP 60069, Avenue Poincar\'e, 59652, Villeneuve d'Asq, France}

\author{Abdelkarim Ouerghi}
\affiliation{Laboratoire de Photonique et de Nanostructures (CNRS-LPN), Route de Nozay, 91460 Marcoussis, France}

\author{Robson Ferreira}
\affiliation{Laboratoire Pierre Aigrain, Ecole Normale Sup\'erieure\\
CNRS (UMR 8551), Universit\'e Pierre \& Marie Curie, Universit\'e Paris Diderot\\
24 rue Lhomond, 75231 Paris Cedex 05, France}

\author{Louis-Anne de Vaulchier}
\email{louis-anne.devaulchier@ens.fr}
\affiliation{Laboratoire Pierre Aigrain, Ecole Normale Sup\'erieure\\
CNRS (UMR 8551), Universit\'e Pierre \& Marie Curie, Universit\'e Paris Diderot\\
24 rue Lhomond, 75231 Paris Cedex 05, France}

\author{Yves Guldner}
\affiliation{Laboratoire Pierre Aigrain, Ecole Normale Sup\'erieure\\
CNRS (UMR 8551), Universit\'e Pierre \& Marie Curie, Universit\'e Paris Diderot\\
24 rue Lhomond, 75231 Paris Cedex 05, France}


\date{\today ~-- v 1.0}


\begin{abstract}
We show that the Landau levels in epitaxial graphene in presence of localized defects are significantly modified compared to those of an ideal system.
We report on magneto-spectroscopy experiments performed on high quality samples. Besides typical interband magneto-optical transitions,
we clearly observe additional transitions that involve perturbed states associated to short-range impurities such as vacancies. Their intensity is found to decrease
with an annealing process and a partial self-healing over time is observed. Calculations of the perturbed
Landau levels by using a delta-like potential show electronic states both between and at the same energies of the Laudau levels of ideal graphene.
The calculated absorption spectra involving all perturbed and unperturbed states are in very good agreement with the experiments.
\end{abstract}

\maketitle


The experimental and theoretical study of graphene, the first 2D crystal ever observed in nature, has become a major issue of condensed
matter physics, both because of its remarkable fundamental physical properties \cite{goerbig2011,castroneto2009} and its promising technological
applications \cite{novoselov2012}. Thermal decomposition of SiC allows to obtain  multilayer epitaxial graphene (MEG) with a low doping level
and high carrier mobility. The graphene layers appear decoupled as MEG essentially displays electronic properties similar to that of an isolated graphene
monolayer. A rotational stacking of the graphene sheets is most frequently proposed to explain this decoupling between adjacent layers \cite{hass2008,deheer2011}.
In fact, C-SiC graphene films are typically comprised of small domains that vary in layer thickness and stacking order \cite{johansson2011}.
The decoupled graphene layers  are quasi neutral with a Fermi level close to the Dirac point with only a few layers close to the interface MEG/SiC being highly doped \cite{sadowski2006}.

The electronic properties of epitaxial multilayer graphene have been widely investigated by infrared magneto-optical spectroscopy \cite{orlita2010,orlita2008}.
The Landau level structure, the mobility of the Dirac fermions as a function of temperature and the nature of the layer stacking  have been accurately determined in MEG
samples containing up to 100 layers. Nevertheless,  it is well-known that lattice defects, either localized (impurities or vacancies) or extended (dislocations, grain boundaries...),
are always present in graphene and that the electronic properties can be significantly modified compared with that of the perfect system \cite{castroneto2009}.
It is therefore important to investigate the magneto-optical signature of the defects in MEG. Surprisingly, hardly any studies about the lattice defects has been
reported using magneto-optical measurements. For example the study of the cyclotron resonance linewidth and the investigation of the dynamical
magneto-conductivity have concluded that the dominant scattering mechanism is very likely due to short-range potentials or to electron-electron interactions \cite{orlita2008,orlita2011}.

In this paper, we present magneto-spectroscopy measurements performed on high-quality MEG samples with intentionally few layers ($\sim$ 8--10)
and high electron mobility of about $2\times 10^5\,\text{cm}^2\text{V}^{-1}\text{s}^{-1}$.
Besides the usual magneto-optical transitions previously observed in MEG corresponding to a single graphene layer (with a characteristic $\sqrt{B}$-dependence)
or to a graphene bilayer (with a quasi linear $B$-dependence), we clearly observe additional transitions, which also display a $\sqrt{B}$-energy dependence.
In the following we show that these transitions involve perturbed states associated to short-range impurities such as C-vacancies. Their intensity is found to decrease with an
annealing process or a few weeks after the growth, indicating a partial self-healing of the layers. To interprete our data we have calculated the perturbed MEG Landau levels in the $\bold{k.p}$
framework by modeling the short-range impurities by delta-like potentials. The important result is that one obtains impurity-related states (i) in between and (ii) at the same
energies as the unperturbed Landau levels of the ideal graphene. We then calculate the magneto-optical interband absorption between all the perturbed and
unpertubed states. The calculated transmission spectra are in good agreement with the experiments.

The investigated MEG films were prepared by thermal decomposition from the C-terminated surface of semi-insulating 4H-SiC substrate \cite{deheer2011,pallecchi2014}.
The SiC substrate was first etched in a hydrogen flux at $1500^\circ\text{C}$ at $200\,\text{mbar}$ for 15 min in order to remove any damage caused by surface
polishing and to form a step-ordered structure on the surface. The graphene layers were grown
in a closed RF induction furnace at a temperature around $1550^\circ\text{C}$ at $10^{-5}\,\text{mbar}$.
Thicknesses of 8--10 layers were intentionally chosen and determined by STEM analysis. Each sample was divided into two parts, one of them being exposed
to pure hydrogen flux at $820^\circ\text{C}$ during 10 minutes in order to check the effect of an H$_2$ annealing \cite{pallecchi2014,pallecchi2012}. Three different sets of samples were investigated
which give similar results. The samples were measured right after the growth and one month
later in order to check the self-healing of the defects. The graphene layers are quasi-neutral with only some layers close to the interface being significantly doped.
To measure the magneto-transmission in the range 5--700$\,\text{meV}$, samples (typically $5\times 5\,\text{mm}^2$) were placed in a $15\,\text{T}$
 superconducting coil at $4\,\text{K}$ and exposed to the radiation of a mercury lamp. The transmitted light was analyzed by a Fourier transform spectrometer,
 using a Si composite bolometer directly below the sample for detection. The samples are partially or fully
 opaque in the energy range 85--210$\,\text{meV}$ because of the phonon-related absorption of the SiC substrate. The transmission at a given magnetic field
 $T(B)$ was normalized by the zero-field transmission $T(0)$.


\begin{figure}
\centering
\includegraphics[width=0.48\textwidth]{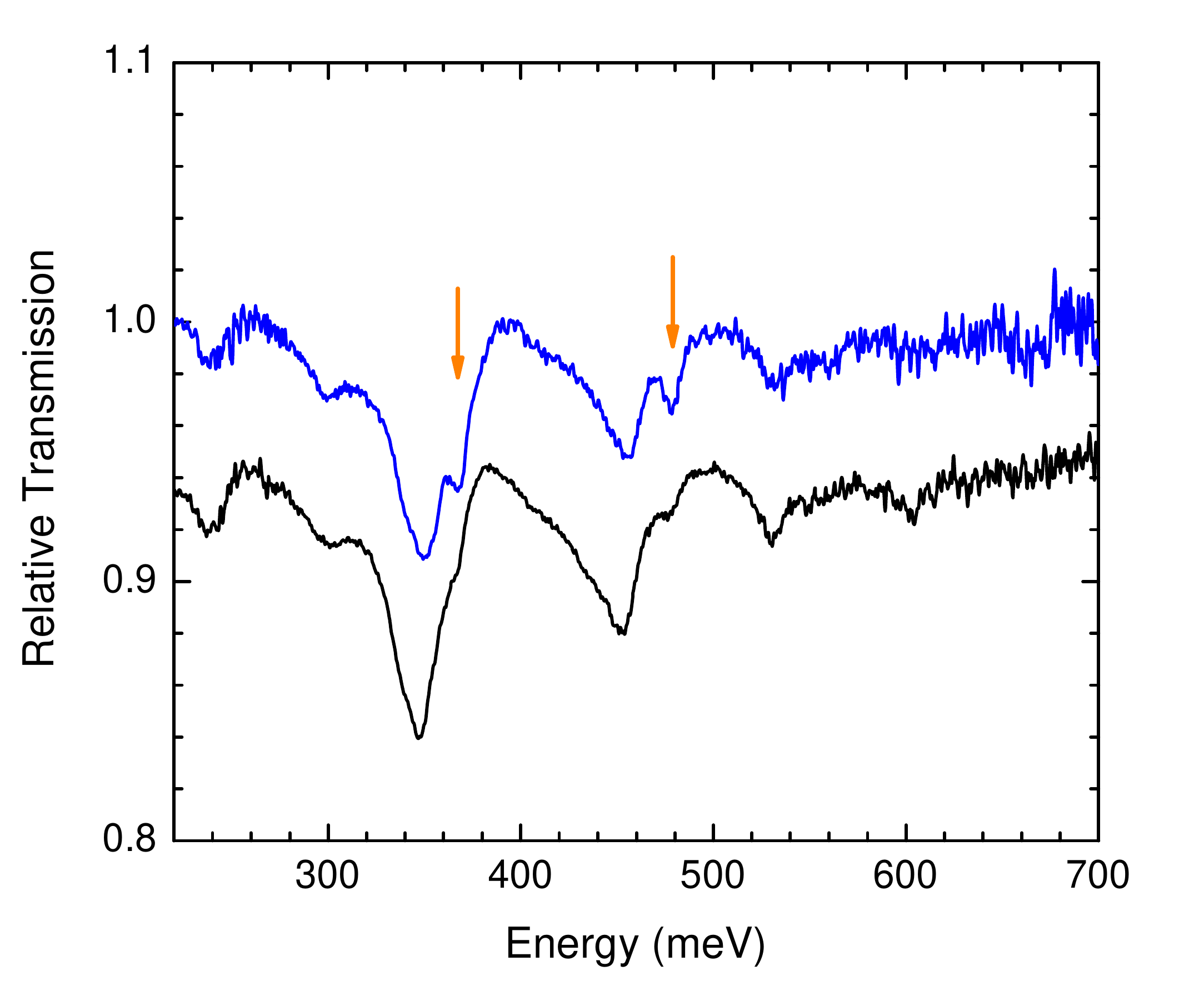}
\caption{Transmission spectra at $B=15\,\text{T}$ for the H$_2$-annealed part (black) and the non-annealed part (blue) of the same sample.
Measurements were done right after the sample growth. The arrows indicate the additional lines due to inter-band transitions between Landau levels and defects states.}
\label{figure1}
\end{figure}

\begin{figure}
\centering
\includegraphics[width=0.47\textwidth]{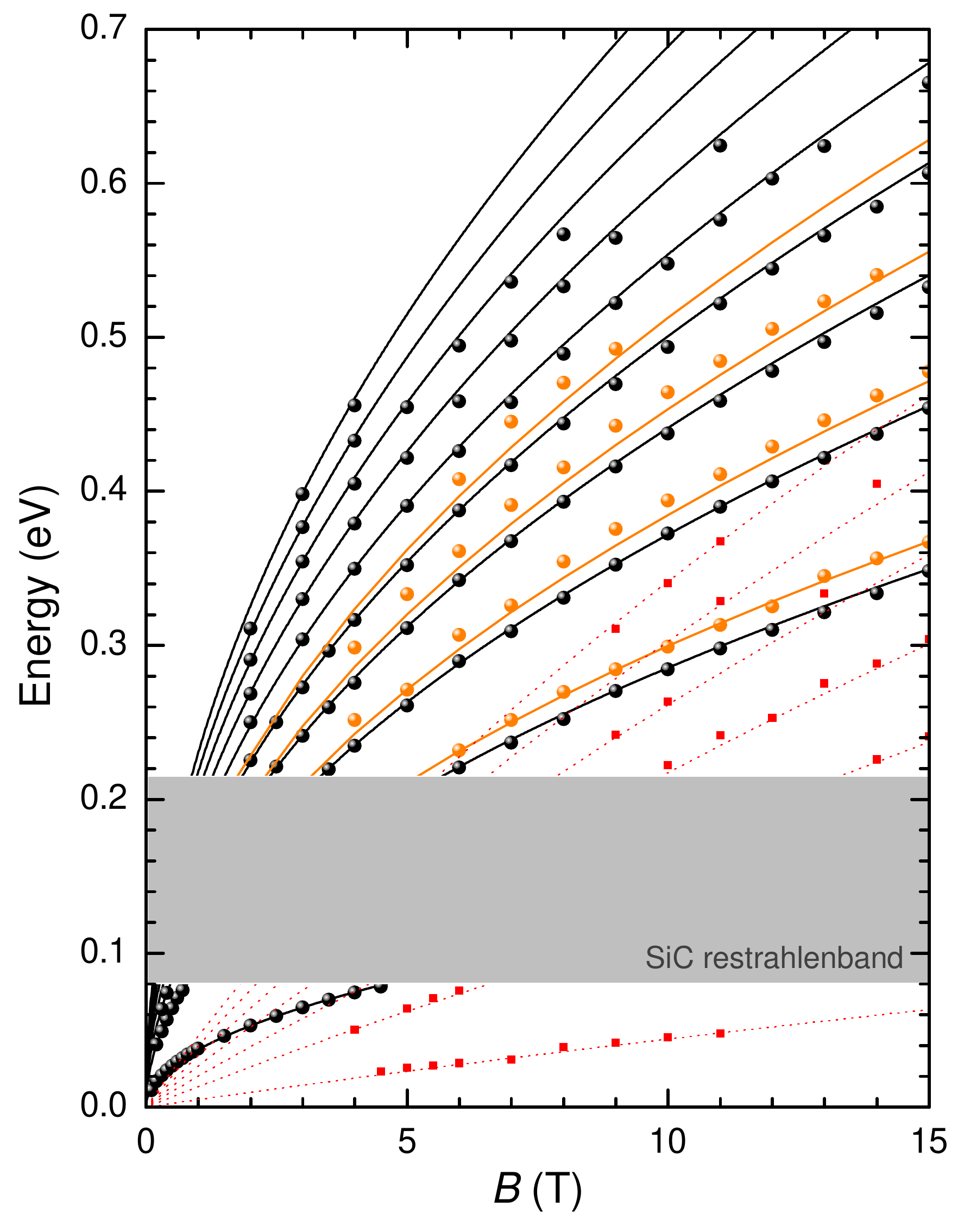}
\caption{Energy position of the observed transmission minima as a function of $B$ (symbols): the dominant absorption lines (black), the defect-induced transitions (orange)
and the bilayer transitions (red) are shown. The solid and dashed lines are the calculated transition energies as described in the text.}
\label{figure2}
\end{figure}

Typical transmission spectra measured at magnetic field  $B=15\,\text{T}$ are shown in Fig.~\ref{figure1} for photon energies above the SiC restralhen
 ($E > 210\,\text{meV}$). The black spectrum is measured on the H$_2$-annealed part of the sample while the blue one corresponds to the non-annealed part
of the same sample. The energy positions of the transmission minima are identical for both spectra but the lines occurring at 365 and $480\,\text{meV}$
(indicated by the arrows) are clearly more pronounced in the blue one. Similar experiments have been performed for magnetic fields in the range $0-15\,\text{T}$
and the position of the transmission minima are shown on Fig.~\ref{figure2}. The dominant absorption lines (shown by the black symbols on Fig.~\ref{figure2})
clearly display the characteristic $\sqrt{B}$-dependence and correspond to transitions between holes and electrons states in a pure graphene monolayer.
Landau levels (LL) of pure graphene in the presence of a perpendicular magnetic field are described by replacing $\bold{k} \rightarrow -i\, \bold{grad} + e\bold{A}/\hbar$ in the lowest
 order ($\bold{k.p}$ approximation) tight-binding hamiltonian, with $\bold{A}$ the potential vector.  This generates the well known LL discrete series
 $E_n=\text{sign}(n) v_F \sqrt{2\hbar e|n|B}$ around the Dirac point at $E = 0$ ($n=n_c > 0$ and $n=n_v < 0$ corresponds respectively to conduction and valence
 band states; $v_F$ is the Fermi velocity). The selection rules for interband absorption within these states in the dipole approximation correspond to $\Delta n = n_c+n_v = \pm 1$.
 The $n=0$ LL can contribute both as initial or final level for the magneto-absorption, depending on the sample doping.  The black solid lines on Fig.~\ref{figure2}
 are the calculated transitions energies for the first ten interband transitions (i.e.: involving LLs $n = -10$ to 10) using $v_F=1.03 \times 10^6\,\text{ms}^{-1}$.
 The agreement is excellent even for the largest transition energies ($600\,\text{meV}$) for which the initial or final LL-energy is $\sim 300\,\text{meV}$
 above and below the Dirac point. The lowest transition $0 \rightarrow 1$ is measured in the energy range 0--85$\,\text{meV}$ below the SiC restralhen. It is noticeable
 that it can be observed for magnetic field as small as $50\,\text{mT}$, which demonstrates the high quality of the MEG samples with a Fermi energy $E_F \leq 10\,\text{meV}$
 and an electron mobility $\mu  > 2\times 10^5\,\text{cm}^2\text{V}^{-1}\text{s}^{-1}$.

We also observed weaker lines which correspond to a nearly linear $B$-dependence as shown by the red symbol on Fig.~\ref{figure2}. The intensity
 of these lines is found to be independent of the hydrogen annealing process (see for instance the two lines at 240 and $300\,\text{meV}$ on Fig.~\ref{figure1}).
 This quasi-linear $B$ dependence, characteristic of massive particles, is associated to graphene bilayers within the MEG. These stacking faults
 in the rotationally ordered MEG are observed in most epitaxial graphene samples \cite{orlita2011b}. It has been reported that the ratio between
 bilayers and monolayers is typically $10\%$. The LL energies of a graphene bilayer can be easily calculated by using a four band model taking
 into account the most relevant coupling constants \cite{koshino2008} and the red dashed lines on Fig.~\ref{figure2} are the calculated energies of the
 five first transitions using $v_F=1.03 \times 10^6\,\text{ms}^{-1}$ and the coupling constant $\gamma_1= 0.39\,\text{eV}$. The agreement with the
 experiments (red symbols) is good.

\begin{figure}
\centering
\includegraphics[width=0.38\textwidth]{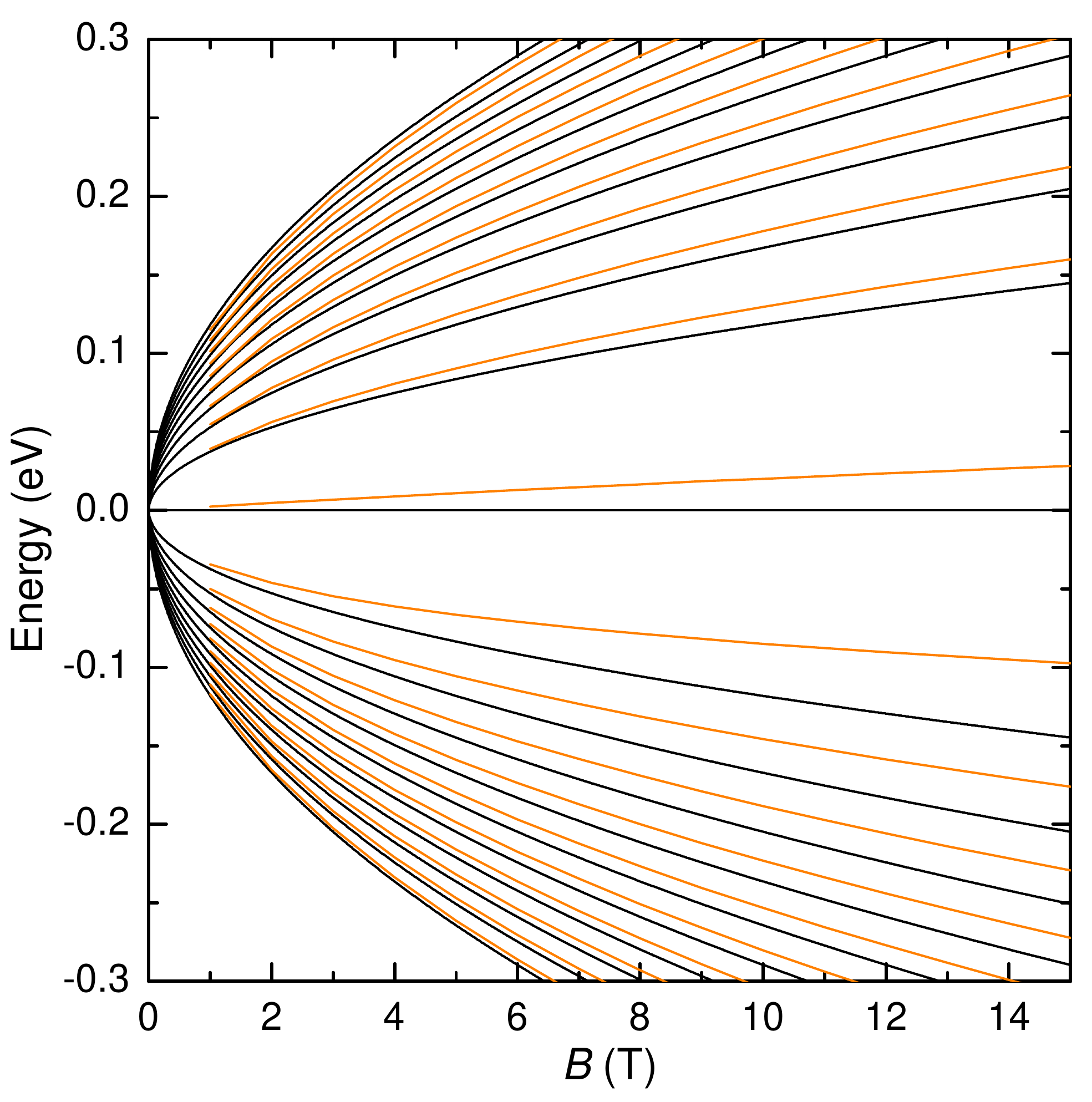}
\caption{Fan chart of perturbed states obtained by diagonalization of $V(r)$ in a truncated basis of unperturbed Landau states ($W_0 = 20\,\text{eV}\,\text{nm}^2$).
Two kinds of states are calculated in the presence of defect: in between (orange) and at the same energies (black) as the unperturbed Landau levels.}
\label{figure3}
\end{figure}

Finally, the major point of this study, additional transmission minima are observed at all magnetic fields with energy shown by the orange
 symbols on Fig.~\ref{figure2}.  They display a  $\sqrt{B}$-dependence and correspond to magneto-optical transitions with energies above that of
 the $n_v \rightarrow n_c$ transitions of ideal graphene (see Fig.~\ref{figure2}). Their intensities are found to be dependent on the annealing process
 (see for instance the lines indicated by the arrows in Fig.~\ref{figure1} which are much stronger for the blue spectrum than for the black one).
 We discuss now the origin of these transitions and we show that they can be associated to short-range impurities (such as C-vacancies) in the graphene layers.
 The effect of defects on the magneto-states has been previously theoretically discussed in the literature \cite{ando1998,shon1998,peres2006}.
 Of particular importance are localized perturbations, owing for instance to a C-vacancy  or one isoelectronic substitution (e.g. one Si atom in place of one C atom)
 on a sublattice A or B site. These have been previously modeled in the $\bold{k.p}$ framework by short-range (delta-like) potentials in real space.
 We have correspondingly considered the perturbation potential $V(r) =W_0 \delta (r)$, where the origin $r=0$ is taken at one site of, say, the A sublattice.
The constant $W_0$, which gives the strength of the localized perturbation, is taken as a parameter to best-fit the experiments.  Finally, we use in the calculations
the symmetric gauge to describe the magnetic field, so that for each Landau level of a given valley there is only one unperturbed Landau state affected by the defect
potential. Namely the one with vanishing angular momentum in the A-component of its spinor (the only states for which the wavefunctions have non-vanishing amplitude at $r =0$).

 \begin{figure}[t]
\centering
\includegraphics[width=0.48\textwidth]{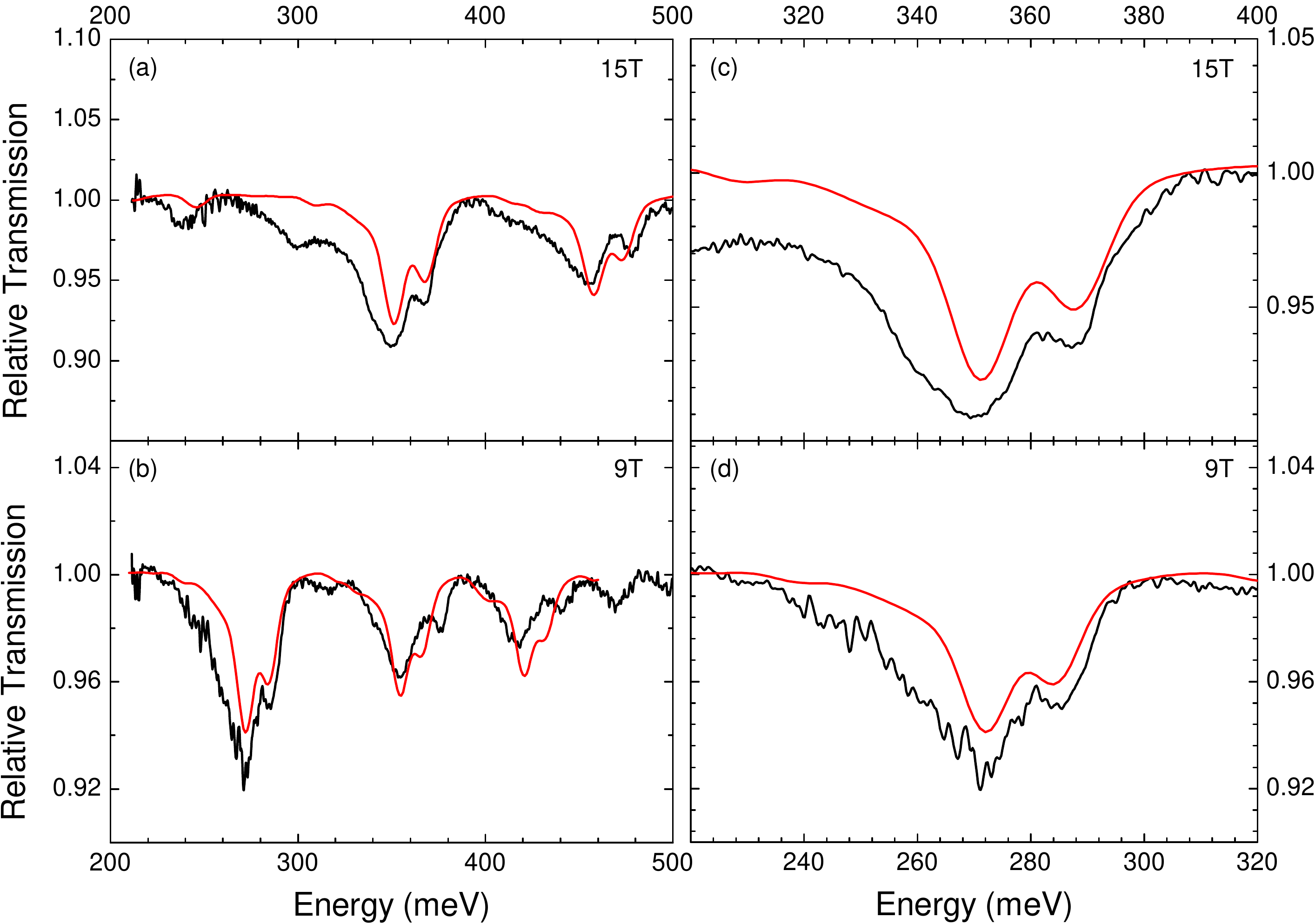}
\caption{Calculated absorption spectra (red) for two magnetic fields: $B = 15\,\text{T}$ ((a) and (c)) and $9\,\text{T}$ ((b) and (d)). Comparison with the experimental
spectra (black) measured in a non-annealed sample right after the growth. (c) and (d) are zooms of (a) and (b) spectra respectively.}
\label{figure4}
\end{figure}

We show in Fig.~\ref{figure3} the fan chart of perturbed states for $W_0 = 20\,\text{eV nm}^2$, obtained after diagonalization of $V(r)$ in a truncated basis of
unperturbed Landau states (a cutoff value $n=10$ is chosen for this calculation, so that 21 LLs are taken into account in the basis).  It is interesting to note
the presence of two kinds of states in the presence of defects: states placed (i) in between and (ii) at the same energies as the unperturbed Landau levels.
The first (in orange in Fig.~\ref{figure3}) is ubiquitous for the perturbation of a general discrete series of levels: the defect introduces new cyclotron orbits with different
energies for the electrons.  The second (in black in Fig.~\ref{figure3}), on the other hand, results from two specificities of a localized perturbation of
Landau states in graphene, i.e., that it effectively mixes states from the same, as well as from different, Dirac cones (while intervalley coupling vanish for extended potentials),
and, additionally, the intra- and inter-valley coupling strengths are identical and equal to $\eta = W_0/ {\lambda_c}^2$, where $\lambda_c$ is the cyclotron radius.
The structure of the spectrum can therefore be understood as follows. At the lowest order in perturbation (i.e., by retaining only the two unperturbed resonant states with same $n \neq 0$
but issued each from a different valley), the eigenvalue equation to solve is: $(E_n + \eta) \openone + \eta \sigma_x = E\,\openone$, with $\openone$ the $2\times 2$ unit matrix,
$\sigma_x$ the real non-diagonal Pauli matrix. The solutions are : $E_\pm = E_n+ \eta \pm | \eta |$. One has thus one state at the same energy $E_n$ as the original Landau level,
and another either below or above it, depending on the sign of $W_0$.

For $n=0$ this same picture does not apply, since the spinors have either A or B components, and states related to the B sublattice remain unaffected by the perturbation.
These lowest order trends obtained for fixed n are also obtained when all (intra- and inter-$n$, for the same or different valleys) couplings are accounted for in the complete diagonalization procedure, resulting in the existence of defect-related states in-between the non-perturbed Landau levels as well as at the energies $E_n$, as shown in Fig.~\ref{figure3}.
Magneto-optical interband absorption involves perturbed as well as unperturbed states.  The measured transmission spectrum is dominated by the interband transitions between unperturbed LLs. In the presence of defects new lines appear, for which the $\Delta n =\pm 1$ selection rules for an ideal graphene layer no longer apply, because of the disorder-induced state mixing.
We obtain that transitions involving only the perturbed states do not contribute to the absorption.  We associate the weaker lines to transitions between unperturbed
and defect-related states.  Indeed, we show in Fig.~\ref{figure4} the calculated absorption probability spectra for two magnetic fields ($B = 9\,\text{T}$ and $15\,\text{T}$),
for electrons initially in unperturbed valence states (LLs with $n \leq 0$) and towards all accessible excited perturbed states (i.e., with $E \geq 0$, the $E=0$ level playing
as for the ideal case the twofold role of initial and final level).  We see that the calculation agrees with the experimental data measured in a non-annealed sample
right after the growth. It is worth stressing a few points related to the theory-experiment comparison in Fig.~\ref{figure4}. Firstly the energy positions of the weak shoulders are mostly
dependent upon the value of the potential strength $W_0$.  Secondly the relative intensities of the main and weaker lines are mostly dependent upon the density of defects.
Indeed, total optical spectrum contains both "LL-to-LL" and "LL-to-defect" contributions.  The intensity of a defect-related contribution is proportional to the number
of defects in the sample, whereas the strength of an intrinsic peak is proportional to the sample surface.  The relative amplitudes between defect-related and intrinsic
contributions are thus proportional to the defect's area density $N_{\text{def}}$, a quantity unknown in our sample and thus considered as a fitting parameter.
Moreover, owing to the fact (discussed above) that some defect-related states have the same energy as the unperturbed Landau orbits, each main line contains both
one LL-to-LL and one LL-to-defect contributions, the latter being proportional to $N_{\text{def}}$. We obtain $N_{\text{def}} = 4.5\times 10^{11}\,\text{cm}^{-2}$
from the best fit shown in Fig.~\ref{figure4} for the non-annealed sample. A similar fit for an H$_2$ annealed sample (see for instance the black spectrum in Fig.~\ref{figure1})
gives $N_\text{def} = 1\times 10^{11}\,\text{cm}^{-2}$. Thirdly in the calculations we replaced the Dirac-lines in the Fermi golden rule by gaussian ones
with energy-independent broadening: $\Gamma_\text{LL-to-LL} = 10\,\text{meV}$ and $\Gamma_\text{LL-to-defect} = 5\,\text{meV}$.  Indeed, we clearly see
in Fig.~\ref{figure4} that the main lines are broader than the weaker ones. The spectra have been calculated in the whole magnetic field range and the position
of the transmission minima associated to the defects is shown by the black and the orange solid lines on Fig.~\ref{figure2}. Again, the agreement with the experimental
results (black and orange symbols) is satisfying.

Let us finally comment on the parameters of the model extracted from the fit in Fig.~\ref{figure4}.  The potential strength is usually written as $W_0 = V_0\, S_{UC}$,
where $S_{UC} = 0.052\,\text{nm}^2$ is the area of the graphene unit cell \cite{ando1998}.  In this case one obtains $V_0 \approx 382\,\text{eV}$.
A short-range potential with large strength has been associated to vacancies in the literature \cite{basko2008,fischer2011}.  Additionally, one finds that
$N_\text{def}/N_C\approx 10^{-4}$ for the non annealed sample and $\approx 2 \times 10^{-5}$ for the annealed sample, where $N_C = 2/S_{UC}$ is the
areal density of carbon atoms.  Note that a similar defect healing is obtained in the non annealed samples measured one month after the growth, indicating
a partial self-healing of the defect even at room temperature. This healing could be explained by a migration at the C-vacancies towards the domain boundaries \cite{cockayne2012,lee2005}.
In any case, the values found for $N_\text{def}/N_C$ corresponds to very diluted concentration
of defects in the graphene layer, consistent with the high electron mobility deduced from the low $B$ cyclotron resonance ($0 \rightarrow 1$ transition) measurements. Moreover the mobility evaluated (following Ref.~\cite{shon1998}) with the disorder parameters is reasonably consistent with the measured one.
Finally, the different broadenings (with $\Gamma_\text{LL-to-LL} \approx 2 \Gamma_\text{LL-to-defect}$) might be due to the fact that the width of the LL-to-LL
transitions results from the convolution of two nearly equally broadened LLs, whereas we do not expect an important broadening for a spatially localized state in the gap between two LLs.

In conclusion, we have observed the magneto-optical signature of short-range impurities in high electron mobility MEG samples.
Interband transitions and magneto-absorption between perturbed and unperturbed states were successfully modeled by using
a delta-like potential in the $\bold{k.p}$ framework. This good agreement demonstrates that the presence of short-range impurities in MEG,
even with a very diluted concentration, must be taken into account for analyzing the magneto-optical data. Such defects are found
to be present with a density of a few $10^{11}\,\text{cm}^{-2}$, even in high electron mobility samples, and give states both between
and at the same energies of the Landau levels of ideal graphene.


\end{document}